\documentclass[9pt, conference]{IEEEtran}
\IEEEoverridecommandlockouts

\usepackage{cite}
\usepackage{amsmath,amssymb,amsfonts}
\usepackage{algorithmic}
\usepackage{graphicx}
\usepackage{textcomp}
\usepackage{xcolor}
\usepackage{multicol, multirow, makecell}
\usepackage{cite}
\usepackage{url}
\usepackage{algorithm}
\usepackage{algorithmic}
\usepackage{booktabs}

\PassOptionsToPackage{hyphens}{url}\usepackage{hyperref}

\def\BibTeX{{\rm B\kern-.05em{\sc i\kern-.025em b}\kern-.08em
    T\kern-.1667em\lower.7ex\hbox{E}\kern-.125emX}}

\renewcommand{\baselinestretch}{0.98}

\begin{document}

\setlength{\abovedisplayskip}{0.5mm}
\setlength{\belowdisplayskip}{0.5mm}

\title{Language-based Audio Moment Retrieval}

\author{\IEEEauthorblockN{Hokuto Munakata}
\IEEEauthorblockA{
LY Corporation\\
hokuto.munakata@lycorp.co.jp}
\and
\IEEEauthorblockN{Taichi Nishimura}
\IEEEauthorblockA{
LY Corporation\\
tainishi@lycorp.co.jp}
\and
\IEEEauthorblockN{Shota Nakada}
\IEEEauthorblockA{
LY Corporation\\
shota.nakada@lycorp.co.jp}
\and
\IEEEauthorblockN{Tatsuya Komatsu}
\IEEEauthorblockA{
LY Corporation\\
tatsuya.komatsu@lycorp.co.jp}
}

\maketitle

\begin{abstract}
In this paper, we propose and design a new task called audio moment retrieval (AMR). Unlike conventional language-based audio retrieval tasks that search for short audio clips from an audio database, AMR aims to predict relevant moments in untrimmed long audio based on a text query. Given the lack of prior work in AMR, we first build a dedicated dataset, Clotho-Moment, consisting of large-scale simulated audio recordings with moment annotations. We then propose a Detection Transformer-based model, named Audio Moment DETR (AM-DETR), as a fundamental framework for AMR tasks. This model captures temporal dependencies within audio features, inspired by similar video moment retrieval tasks, thus surpassing conventional clip-level audio retrieval methods. Additionally, we provide manually annotated datasets to properly measure the effectiveness and robustness of our methods on real data. Experimental results show that AM-DETR, trained with Clotho-Moment, outperforms a baseline model that applies a clip-level audio retrieval method with a sliding window on all metrics, particularly improving Recall1@0.7 by 9.00 points.
Our datasets and code are publicly available in \url{https://h-munakata.github.io/Language-based-Audio-Moment-Retrieval}.
\end{abstract}


\section{Introduction}
Language-based audio retrieval, referred to simply as audio retrieval, is to search for desired audio from an audio database using a natural language query. Researchers pay attention to this technology because of its wide range of applications, such as a retrieval system of sound archives and sound effects~\cite{koepke2022audio}. With an introduction of large-scale audio-text datasets~\cite{drossos2020clotho, kim2019audiocaps, mei2024wavcaps}, audio retrieval methods that process audio and text in an embedding space have been investigated~\cite{elizalde2024natural, yusong2023laionclap, saeed2021contrastive, niizumi24_interspeech, Primus2023}.

Because previous works use audio-text datasets consisting of short audio (5 to 30 seconds), they proposed a method for retrieving short audio from natural languages. Inspired by cross-modal retrieval in the vision domain~\cite{radford2021learning}, the mainstream approach is to employ contrastive learning with audio-text pairs~\cite{elizalde2024natural, yusong2023laionclap, saeed2021contrastive, niizumi24_interspeech, Primus2023}.
This model maps both the audio clip and text query into a shared embedding space, retrieving the relevant audio by calculating the similarity between the embeddings of the text query and the audio in the database.

\begin{figure}
    \centering
    \includegraphics[width=0.99\linewidth]{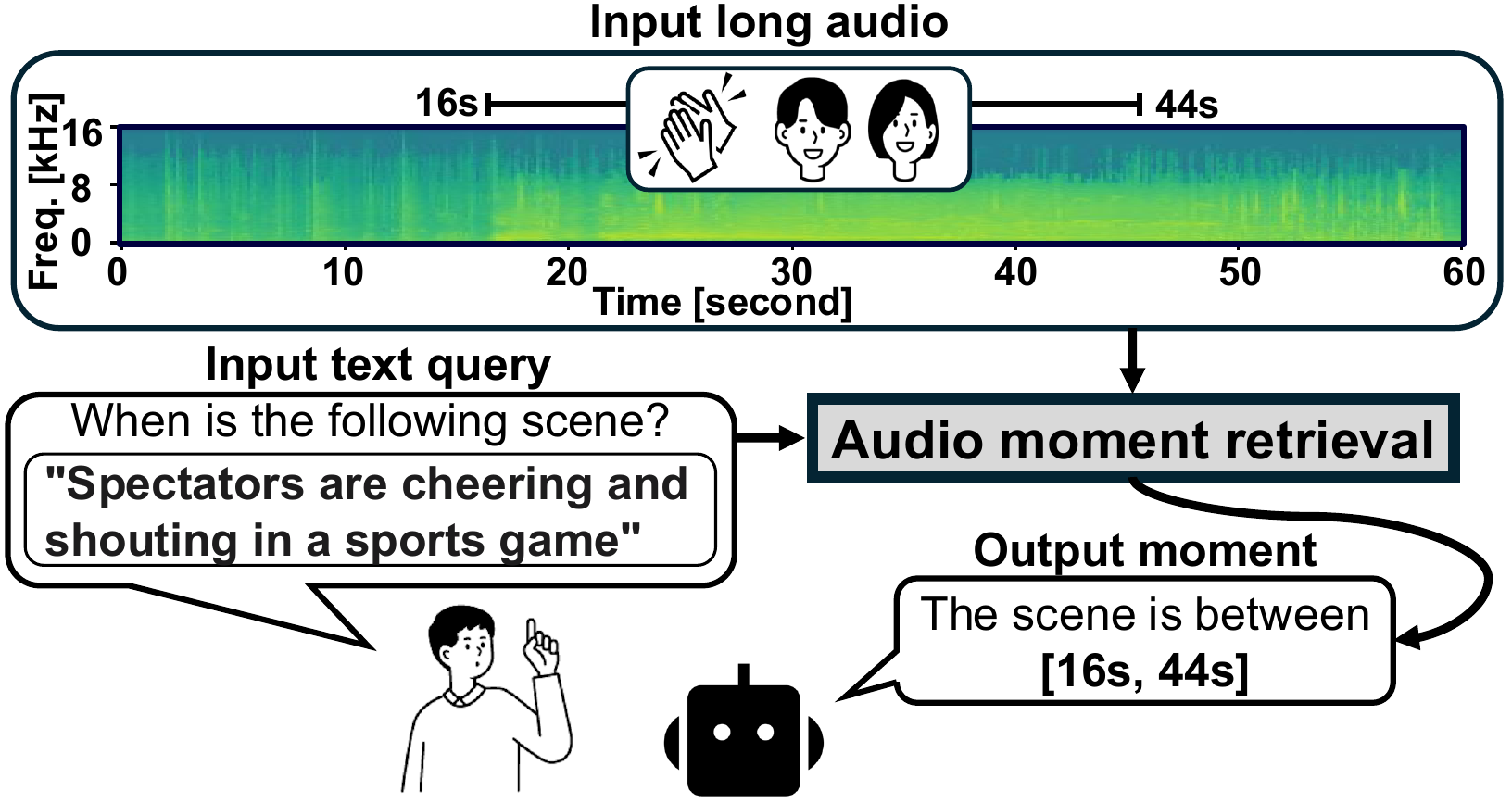}
    \vspace{-8mm}
    \caption{An overview of audio moment retrieval. Given a long audio and text query, the system retrieves relevant moments as start and end time stamps.}
    \label{fig:task_overview}
    \vspace{-4mm}
\end{figure}

Although existing audio retrieval models assume that the audio is trimmed as a short clip, there is a great demand for retrieving specific time segments from untrimmed long audio.
Potential applications include, for example, highlight moment detection for sports broadcasts and criminal moment detection for surveillance systems.
These applications motivate us to retrieve moments in the audio using natural language queries.
Figure~\ref{fig:task_overview} shows the retrieval system we envision to implement the application.
Given a query ``Spectators are cheering and shouting in a sports game'' along with long audio, the model is expected to retrieve \textit{audio moments} as start and end timestamps (16 to 44 seconds).
A straightforward solution to building such moment retrievers is to utilize existing audio retrieval methods in three steps: dividing the long audio into multiple clips, forwarding them into the audio retrieval models, and calculating the similarity between the text query and the audio clip embeddings in the shared embedding space. However, this approach is sub-optimal because it processes the audio clip sequence independently, failing to capture the temporal dependencies between the clips.

To address this limitation, we can draw inspiration from the field of the video domain, where modeling temporal dependencies is well explored. Video moment retrieval (VMR) is a similar task that retrieves specific moments from a long video~\cite{anne2017localizing, gao2017tall, lei2020tvr, lei2021detecting, moon2023query, moon2023correlation, xiao2024bridging}.
For this task, DEtection TRansformer (DETR)-based models~\cite{carion2020end} show high retrieval performance~\cite{lei2021detecting, moon2023query, moon2023correlation}.
These models learn not only temporal dependencies inside the video frames but also cross-modal similarity between frames and text query, resulting in correct moment prediction.
Various methods have been proposed in VMR and we think the ideas of these methods are helpful to our target system.

\begin{figure*}
    \centering
    \includegraphics[width=0.95\linewidth]{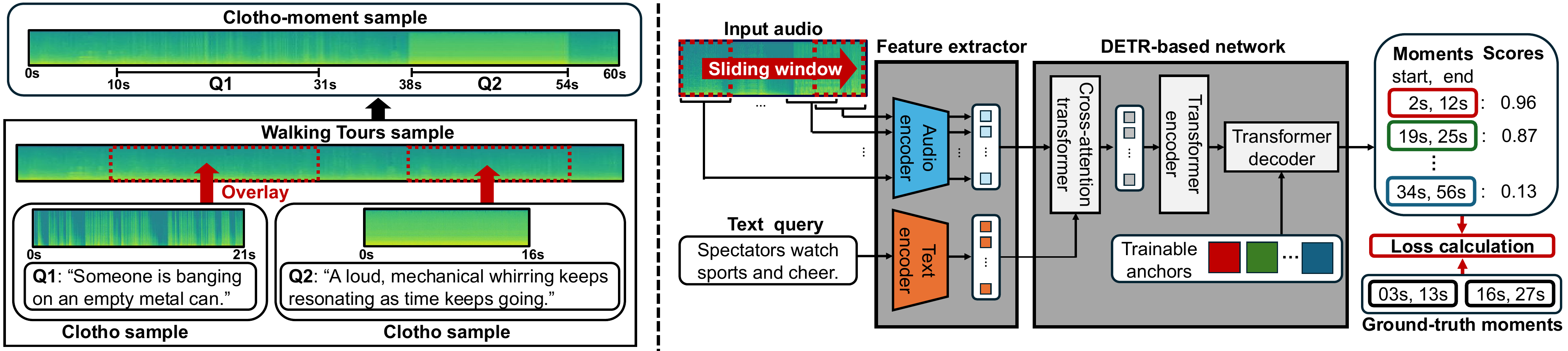}
    \vspace{-3mm}
    \caption{The left figure describes Clotho-Moment. By overlaying Clotho on Walking Tours, audio and text query-audio moment pairs are generated.
    The right side of the figure describes the architecture of the proposed Audio Moment DETR. This model first extracts embedding by audio/text encoder and then the DETR-based network transforms the embedding to pairs of the audio moment and its confidence score.}
    \vspace{-3mm}
    \label{fig:model_arch}
\end{figure*}

In this paper, we propose and design a new task called \textit{audio moment retrieval} (AMR) inspired by the methodologies and successes seen in VMR.
Given the lack of prior work in AMR, we first build a dataset dedicated to AMR, namely Clotho-Moment, which consists of large-scale simulated audio recordings with moment annotations.
Then, we propose DETR-based AMR models named Audio Moment DETR as a fundamental framework for the AMR tasks, moving beyond the conventional audio \textit{clip} retrieval.
Additionally, we provide manually annotated datasets, allowing us to properly measure the effectiveness and robustness of the methods on real data.
We hope that our research shows a new direction of audio retrieval and pave the way for future research in this area.

\section{Difference to Existing Tasks}
\label{sec: diff}
AMR has three similar tasks in audio and vision domains: sound event detection (SED), target sound extraction (TSE), and video moment retrieval (VMR). In this section, we emphasize the novelty of AMR by enumerating the difference between AMR and these tasks.
\\
\textbf{Sound event detection (SED)} is the task of predicting both the sound event class labels and the event time boundaries from the input audio~\cite{cai2003highlight, cakir2017convolutional, komatsu2016acoustic,imoto2020sound,miyazaki2020convolution,komatsu2020scene, mesaros2021sound, li2023ast, shao2024fine}.
SED assumes the target labels are predetermined and each label corresponds to a single event.
In contrast, AMR assumes an open vocabulary for queries and thus can handle queries related to multiple events.
Hence, AMR can be seen as zero-shot SED in an open-vocabulary setting.
\\
\textbf{Target sound extraction (TSE)} is the task of extracting sound sources relevant to the text query from the input mixture~\cite{liu22w_interspeech, li2023target}.
The difference between AMR and TSE lies in the target output.
AMR aims to identify the relevant moment, while TSE aims extract the relevant separated source signal.
Therefore, even if we obtain separated signals with TSE, we still need to perform AMR to retrieve the desired moments.
\\
\textbf{Video Moment Retrieval (VMR)} is a task in which the input and output are the same as AMR, except that VMR focuses on visual data instead of audio~\cite{anne2017localizing, gao2017tall, lei2020tvr, lei2021detecting, moon2023query, moon2023correlation, xiao2024bridging}.
Although there are VMR models that utilize audio features extracted from video~\cite{xiao2024bridging}, audio features are only used as auxiliary information for the video frame.

\section{Dataset for Audio Moment Retrieval}
We construct datasets to retrieve audio moments relevant to the text query from untrimmed audio data.
The audio we assume for AMR is one minute long and includes some specific scenes that can be represented by text, such as audio of a sports broadcast including a scene in which the goal is scored and people cheering. Based on these assumptions, we create two types of datasets.

\subsection{Clotho-Moment}
We propose Clotho-Moment, a large-scale simulated AMR dataset.
The dataset is generated based on the simulation by leveraging existing audio-text data, which does not require additional annotation.
Clotho-Moment is generated from two datasets, Clotho~\cite{drossos2020clotho} and Walking Tours~\cite{venkataramanan2023imagenet}.
These datasets are selected because Clotho samples contain specific scenes that we assume to be the target of AMR, and most parts of Walking Tours samples do not.
\textbf{Clotho} consists of various manually annotated 15 to 30-second audio clip data collected on Freesound.
Each audio sample has five captions comprising 8 to 20 words.
This dataset has development, validation, and evaluation splits and each split contains 3839, 1045, and 1045 audio samples.
\textbf{Walking Tours} contains ten videos over an hour long recorded in various cities.
We use only the audio in the video.
Following Clotho, we split this dataset into development, validation, and evaluation splits in a 7:1:2 ratio.

\textbf{Clotho-Moment} simulates long recorded audio in the city including some scenes contained in Clotho.
As shown on the left side of Figure~\ref{fig:model_arch}, this dataset is generated with Clotho and Walking Tours audio samples as foreground and background, respectively.
Figure~\ref{fig:model_arch}.
Inspired by the simulation method proposed for the speaker diarization~\cite{fujita2019end}, the audio and pairs of text query and moments are generated by overlaying Clotho samples on Walking Tours samples at randomly sampled intervals as shown in Algorithm 1.
Note that our method can only generate a single audio moment per query and each scene is not overlapped.

We make several adjustments to make the data more realistic.
To simulate scenes with various volume levels, the signal power of the sampled audio of Clotho and Walking Tours is randomly weighted at $\pm5$ dB, and at $-20\pm5$ dB, respectively.
To make audio moments more accurate, we removed the silent onsets and offsets of Clotho clips that were 20 dB lower than the overall signal power.
We set hyperparameter $\lambda$ controlling the average interval length to $30$ second.
To reduce mismatch with our manually annotated dataset created described below, Walking Tours samples are divided into one-minute segments in one-second intervals in advance.
Finally, we generated 32694, 4918, and 6649 samples for development, validation, and evaluation split.

\newcommand{\COMMENTOUT}[2][.0\linewidth]{%
  \hfill\makebox[#1][r]{ \scriptsize{// #2}}%
}

\begin{figure}[t]
\vspace{-6mm}
  \begin{algorithm}[H]
    \caption{Generating Clotho-Moment}
    \begin{algorithmic}[1]
    \item[\textbf{Input:}] $\mathcal{D}_{\rm Clotho}$ \COMMENTOUT{Set of of query-foreground audio pairs} \\
    \hspace*{1.2em} ${\mathcal{D}_{\rm WalkingTours}}$ \COMMENTOUT{Set of background audio} 
    \item[\textbf{Output:}] $\mathbf{x}$ \COMMENTOUT{Generated audio} \\
    \hspace*{1.8em} $Y=\{(q_1, t_{\rm 1, start}, t_{\rm 1, end} ),...\}$ \COMMENTOUT{Set of query-moment pairs} \\
    \STATE $Y \leftarrow \emptyset,~t=0$ \COMMENTOUT{Initialize} 
    \STATE $\mathbf{x}\sim \mathcal{D}_{\rm WalkingTours}$ \COMMENTOUT{Sample background audio}
    \WHILE{}
        \STATE $d \sim \frac{1}{\beta}{\rm exp}(-\frac{d}{\beta}$)\COMMENTOUT{Sample interval}
        \STATE ${\mathbf{x}_{\rm moment}, q}  \sim \mathcal{D}_{\rm Clotho}$ \COMMENTOUT{Sample foreground audio and its query}
        \STATE $t_{\rm start} \leftarrow t+d$
        \STATE $t_{\rm end} \leftarrow t_{\rm start}+{\rm duration}(\mathbf{x}_{\rm moment})$
        \STATE \textbf{if} $t_{\rm end} > {\rm duration}(\mathbf{x})$ \textbf{then} \textbf{break}
        \STATE $\mathbf{x}[t_{\rm start}:t_{\rm end}]=  \mathbf{x}[t_{\rm start}:t_{\rm end}]+  \mathbf{x}_{\rm moment}$ \COMMENTOUT{Overlay}
        \STATE $Y \leftarrow Y~|~\{(q, t_{\rm start}, t_{\rm end})\}$ \COMMENTOUT {Update sets}
        \STATE $t \leftarrow t_{\rm end}$
    \ENDWHILE
    \RETURN $\mathbf{x}, Y$
    \end{algorithmic}
  \end{algorithm}
  \vspace{-9mm}
\end{figure}

\subsection{Real Data for Evaluation}
We also create a small-scale manually annotated AMR dataset to evaluate the retrieval performance in more realistic environments.
As a relatively long video data collection containing various scenes, we chose \textbf{UnAV-100}, an untrimmed YouTube video dataset~\cite{geng2023dense}, and extracted audio from the video.
We selected 77 videos and annotated 100 new text queries and audio moments. 
To make the evaluation easy, each query corresponds to a single audio moment. 
The audio duration is from 45 to 60 seconds and the resolution of the moments is one second.
The captions consist of 6.3 words on average.

\section{Audio moment retrieval}
In this section, we define audio moment retrieval (AMR).
To tackle AMR, we propose Audio Moment DETR (AM-DETR) inspired by the VMR models~\cite{lei2021detecting, moon2023query, moon2023correlation}.
By training with Clotho-Moment, AM-DETR enables to capture of both temporal dependencies inside the audio features and cross-modal similarity between audio and text.

\subsection{Task Formulation}
We define audio moment retrieval as the task of predicting audio moments corresponding to a given text query. Let the input audio be denoted as $\mathbf{x}$, the text query as $q$, and the output audio moments as $y$. Here, $y$ represents the audio moment relevant to the query and may consist of multiple moments: $y = (y_1, y_2, \ldots, y_N)$. Each moment is represented as a tuple of start and end times, $y_n = (t_{n,\text{start}}, t_{n,\text{end}})$. This task can be viewed as a mapping problem where, given the input audio $\mathbf{x}$ and the text query $q$, we aim to find the relevant audio moments $y$. Formally, this can be expressed as $y \simeq f_\text{AMR}(\mathbf{x}, q)$.
The key challenge is how to model the function $f_\text{AMR}$.

\subsection{Audio Moment DETR}
We propose Audio Moment DETR (AM-DETR), an AMR model inspired by the model proposed for VMR~\cite{lei2021detecting, moon2023query, moon2023correlation}.
As shown on the right side of Figure~\ref{fig:model_arch}, AM-DETR consists of a feature extractor and a DETR-based network the same as the VMR models.
The key part of AM-DETR is the DETR-based network~\cite{lei2021detecting, moon2023query, moon2023correlation} that captures the temporal dependency between audio features, such as temporal changes in the audio features.
Specifically, the architecture of AM-DETR is based on QD-DETR, which is simple but achieves high retrieval performance for VMR by capturing both the temporal dependency between video frames and cross-modal dependency between frames and text query~\cite{moon2023query}.
\\
\textbf{(1) Feature extractor}.
Feature extractor consists of audio/text encoder.
Given an input audio $\mathbf{x}$ and a text query $q$, each encoder them into embeddings as follows:
\begin{align}
\mathbf{z}^{\text{(T)}} = {\rm TEnc}(q),~
\mathbf{z}^{\text{(A)}} = {\rm AEnc}(\mathbf{x}),
\end{align}
where $\mathbf{z}^{\text{(T)}} \in \mathbb{R}^{D \times L_t}$ and $\mathbf{z}^{\text{(A)}} \in \mathbb{R}^{D \times L'_a}$ represent the $D$-dimensions linguistic and acoustic embeddings with lengths $L_t$ and $L'_a$.
To prompt the following DETR-based network to capture the similarity between the audio and text embedding, we employ pre-trained audio/text encoders based on contrastive learning.
Since the audio encoder is trained with only short clips through contrastive learning, the similarity between the embeddings of long audio and text cannot be measured correctly.
To address this issue, we divide $\mathbf{x}$ into $L_a$ short clips $\mathbf{x}_1, \ldots, \mathbf{x}_{L_a}$ using the sliding window in advance. 
We encode the short clips and pool the embedding along the temporal axis independently as follows:
\begin{align}
\mathbf{z}^{\rm (A)} = \left[\text{Pool}({\rm AEnc}(\mathbf{x}_1)), \ldots, \text{Pool}({\rm AEnc}(\mathbf{x}_{L_a}))\right] \in \mathbb{R}^{D\times L_a}.
\end{align}
We set the hop length of the sliding window to 1 in our experiment.
\\
\textbf{(2) DETR-based network}.
First, this network transforms $\mathbf{z}^{\rm (A)}$ and $\mathbf{z}^{\rm (T)}$ by Cross-attention Transformer to sequential latent feature that captures cross-modal dependency by measuring similarity between audio and text embeddings as follows:
\begin{align}
\mathbf{z} = {\rm CrossAttentionTransformerEnc}(\mathbf{z}^{\rm (A)}, \mathbf{z}^{\rm (T)}).
\end{align}
Furthermore, to capture the temporal dependency between the latent features, Transformer encoder transforms $\mathbf{z}$ leveraging the self-attention mechanism as follows:
\begin{align}
\mathbf{z}' = {\rm TransformerEnc}(\mathbf{z}).
\end{align}
Finally, Transformer decoder transforms $\mathbf{z}'$ to $K$ candidate audio moments $\hat{y}_1,...,\hat{y}_K$ and their confidence scores $c_1,...,c_K\in [0,1]$.
The confidence score represents how plausible the corresponding predicted moment is.
Here, $K$ is the hyperparameter controlling the number of outputs that assumes sufficiently larger than the number of the ground truth moment.
To output $K$ candidates, the decoder also takes $K$ trainable anchors, which are parameters of the neural network controlling the number of output ~\cite{carion2020end} in addition to $\mathbf{z}'$.
Finally, the decoder is formulated as follows:
\begin{align}
(\hat{y}, c) = {\rm TransformerDec}(\mathbf{z'}),
\end{align}
where $\hat{y}=(\hat{y}_1,...,\hat{y}_K)$ and $c=(c_1,...,c_K)$ is a tuple of predicted moments and confidence scores, respectively.
Note that, moments are output as the center and width of the moment relative to the length of the audio because the model is implemented to mimic object detection.
Since the number of ground-truth audio moments $N$ is unknown, we select plausible audio moments from the $K$ candidates by comparing the confidence scores with a threshold.
\\
\textbf{(3) Loss function}.
To output correct audio moments and confidence scores, AM-DETR is trained to minimize two types of losses, moment loss $\mathcal{L}_{\rm moment}$ and score loss $\mathcal{L}_{\rm score}$.
The moment loss consists of $L_1$ loss and generalized Intersection over Union (gIoU) loss~\cite{rezatofighi2019generalized}.
$L_1$ loss measures the error of the center coordinate and width of the predicted audio moment as follows:
\begin{align}
\mathcal{L}_{L_1}(\hat{y}_k, y_n) = \frac{1}{2}\left\|(\hat{t}_{\rm k, start} + \hat{t}_{\rm k, end}) - (t_{\rm n, start} + t_{\rm n, end})\right\|_{L_1} \nonumber\\
+ \left\|(\hat{t}_{\rm k, end} - \hat{t}_{\rm k, start}) - (t_{\rm n, end} - t_{\rm n, start})\right\|_{L_1},
\end{align}
where the first and second terms represent errors of the center coordinate and width, respectively.
The gIoU loss $\mathcal{L}_{\mathrm{gIoU}}(\hat{y}, y)$ represents the negative value of IoU between $\hat{y}$ and $y$ and is used for direct optimization for the IoU.
Finally, the moment loss is obtained by the weighted sum of these losses as follows:
\begin{align}
\label{eq:moment_loss}
    \mathcal{L}_{\rm moment}(\hat{y}_k, y_n) = 
    \lambda_{L_1} \mathcal{L}_{L_1}(\hat{y}_k, y_n)
    + \lambda_{\mathrm{gIoU}}\mathcal{L}_{\mathrm{gIoU}}(\hat{y}_k, y_n).
\end{align}

To predict the confidence scores correctly, the model learns to identify whether each output candidate has a corresponding ground truth using cross-entropy loss.
Here we introduce a new index $\phi=(\phi(1),...,\phi(K))$ for $K$ output candidates and assume that $\phi(1),...,\phi(N)$-th candidates has corresponding ground truth.
Under this assumption, the score loss is obtained as follows:
\begin{align}
\label{eq:ce}
    \mathcal{L}_{\rm score}(c, \phi) = 
    \sum_{i=1}^N
    \left(- \log c_{\phi(i)}\right)
    + \sum_{i=N+1}^K \left(-\log (1-c_{\phi(i)})\right),
\end{align}
where the first term acts to increase the confidence score of moments with a corresponding ground truth, and the second term acts to decrease the score of moments without a corresponding ground truth.

Since the correct correspondence between $K$ candidates and $N$ ground truths is unknown, we determine the optimal correspondence minimizing the following matching loss as follows: 
\begin{align}
\label{eq: permloss}
    \mathcal{L}_{\rm match}(\hat{y},y, c, \phi) = 
    \sum_{n=1}^N
    \left(
    - c_{\phi(n)}
    + \mathcal{L}_{\rm moment}(\hat{y}_{\phi(n)}, y_n).
    \right)
\end{align}
Here, it is empirically known that using $c_{\phi(n)}$ not but $\log c_{\phi(n)}$ improves the performance in Eq~\eqref{eq: permloss}~\cite{carion2020end}.
We denote the optimal index as $\hat{\phi} = \underset{\phi \in \mathcal{P}}{\mathrm{argmin}}~\mathcal{L}_{\rm match}(\hat{y},y, c, \phi)$ where $\mathcal{P}$ is a set of every $K!$ permutation.
Finally, the overall loss is obtained by the weighted sum of the moment loss and the CE loss as follows:
\begin{align}
\label{eq: overallloss}
    \mathcal{L} = 
    \lambda_{\rm score}\mathcal{L}_{\rm score}(c, \hat{\phi}) + \sum_{n=1}^N 
    \mathcal{L}_{\rm moment}(\hat{y}_{\hat{\phi}(n)}, y_n).
\end{align}

\begin{table*}[t]
    \centering
    \caption{Retrieval performance on each dataset. For all metrics, higher is better. }
    \vspace{-3mm}
    \label{tab: amr_score}
    \scalebox{0.80}{
    \begin{tabular}{lccc|rrrrr|rrrrr|rrrrr}
    \toprule
    \multirow{4}{*}{Method} & \multirow{4}{*}{\makecell{Feature\\extractor}} & \multirow{4}{*}
    {\makecell{Contrastive\\learning}} &\multirow{4}{*}{\makecell{Window\\length}} & \multicolumn{5}{c|}{Clotho-Moment (eval)} & \multicolumn{5}{c|}{UnAV-100 subset} & \multicolumn{5}{c}{TUT-Sound Events 2017} \\
    & & & & \multicolumn{2}{c}{R1} & \multicolumn{3}{c|}{mAP} 
    & \multicolumn{2}{c}{R1} & \multicolumn{3}{c|}{mAP} 
    & \multicolumn{2}{c}{R1} & \multicolumn{3}{c}{mAP} \\
    \cmidrule(lr){5-6} \cmidrule(lr){7-9} \cmidrule(lr){10-11} \cmidrule(lr){12-14}
    \cmidrule(lr){15-16} \cmidrule(lr){17-19}
    & & & & @$0.5$ & @$0.7$ & @$0.5$ & @$0.75$ & avg 
    & @$0.5$ & @$0.7$ & @$0.5$ & @$0.75$ & avg 
    & @$0.5$ & @$0.7$ & @$0.5$ & @$0.75$ & avg \\
    \midrule
    Baseline & CLAP & \checkmark & 1.0
    & 20.92 & 14.65 & 22.12 & 13.56 & 13.61
    & 18.00 &  5.00 & 19.50 &  6.00 & 7.45
    &  4.81 &  1.92 &  3.85 &  0.96 & 1.59\\
    
    Baseline & CLAP & \checkmark & 4.0
    & 44.79 & 30.89 & 46.83 & 28.08 & 27.19
    & 53.00 & 24.00 & 55.00 & 23.00 & 25.30
    &  4.81 &  0.96 &  4.73 &  1.48 & 1.93 \\
    
    Baseline & CLAP & \checkmark & 7.0
    & 62.11  & 48.05  & 62.87  & 41.99  & 38.11
    & 55.00 & 32.00 & 57.00 & 24.50 & 29.00
    &  7.69 & 3.85 & 7.39 & 1.68 & 3.07 \\
    
    \midrule
    
    AM-DETR & VR & -  & 1.0
    & 49.90  & 30.98  & 65.19  & 28.42  & 32.92
    & 28.00 & 12.00 & 43.05 & 12.14 & 17.71
    & 6.73 & 3.85 & 11.63 & 2.22 & 4.42 \\

    AM-DETR & VR & \checkmark & 1.0
    & \textbf{88.25}  & \textbf{82.66}  & 91.21  & 82.01  & \textbf{77.09}
    & 58.00 & \textbf{41.00} & 64.31 & \textbf{41.06} & \textbf{40.82}
    & \textbf{15.38} & \textbf{6.73} & 15.06 & 4.09 & 5.55 \\

    AM-DETR & CLAP & \checkmark &  1.0
    & 87.50  & 81.86  & \textbf{91.39}  & \textbf{82.51}  & 76.66 
    & \textbf{61.00} & 39.00 & \textbf{67.08} & 36.42 & 40.30
    & 14.42 & 5.77 & \textbf{15.72} & \textbf{4.38} & \textbf{6.23}  \\
    \bottomrule
    \end{tabular}
    }
\end{table*}

\section{Experiment}
We evaluated the proposed AMR model trained with Clotho-Moment and a baseline.
Our experiments were conducted using the open-source library called Lighthouse\footnote{\url{https://github.com/line/lighthouse}}~\cite{nishimura2024lighthouse}, and the datasets are also publicly available.

\subsection{Model Configuration}
\noindent
\textbf{Proposed model}. 
To investigate the impact of the feature extractor, we used two feature extractors, CLAP~\cite{elizalde2024natural} and VR~\cite{munakata2024dcase}.
VR is also a conventional audio retrieval model consisting of VAST~\cite{chen2024vast} and RoBERTa~\cite{liu2019roberta}.
To confirm whether pre-training of encoders based on contrast learning is effective in AMR, we also used VAST and RoBERTa without contrastive learning as the feature extractors.
For comparison, we used the sliding window with 1, 4, and 7-second window lengths.
The hyperparameter settings of the DETR-based network followed~\cite{moon2023query}. 
Both Transformer encoder and decoder consisted of two stacks of attention layers with 256 units and eight heads, and the dimension of the feed-forward layers was set to 1024. 
We used AdamW optimizer~\cite{loshchilov2017decoupled} with a learning rate of \(1 \times 10^{-4}\) and trained the models for 100 epochs with a batch size of 32. 
\\
\textbf{Baseline}. 
To measure how much AM-DETR trained with Clotho-Moment improves AMR performance by capturing time dependence, we prepared a baseline model using an existing speech retrieval model without training for AMR. 
The baseline model is a combination of the existing audio retrieval model~\cite{elizalde2024natural} with the sliding window.
To obtain audio moments, the baseline measures the similarities between audio and text embeddings for each window frame and then binarizes the similarities by comparing them to a threshold.
The confidence score is defined as the average similarity within the moment. 
To improve performance, we applied a median filter to the resulting binary sequence.
The threshold for the binarization and the length of median filters were tuned with the validation split of Clotho-Moment.

\subsection{Evaluation}
We measured Recall1 (R1) and mean average precision (mAP) which are widely used metrics in VMR~\cite{lei2021detecting}.
Both metrics are based on whether the IoU between the ground truth and predicted moments exceeds threshold $\theta$.
R1 is an evaluation metric for a single audio moment using the candidate with the highest confidence score.
In contrast, mAP is an evaluation metric considering multiple audio moments considering all ground truth and the corresponding candidates.
We also measure average mAP with multiple $\theta$ ranging from 0.5 to 0.95 in 0.05 intervals.

\subsection{Results}
\begin{figure}[t]
    \centering
    \begin{minipage}[b]{0.49\columnwidth}
        \centering
        \includegraphics[width=0.99\columnwidth]{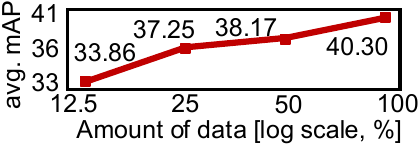}
        \label{fig:win_len_vs_score}
    \end{minipage}
    \begin{minipage}[b]{0.49\columnwidth}
        \centering
        \includegraphics[width=0.99\columnwidth]{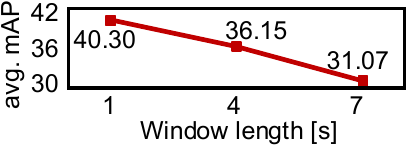}
        \label{fig:data_amount_vs_score}
    \end{minipage}
    \vspace{-6mm}
    \caption{Comparative study for the window length and the amount of data. The vertical axes for both figures represent average mAP on UnAV-100 subset.}
    \label{fig:hparam_vs_score}
    \vspace{-3mm}
\end{figure}

\noindent
\textbf{Proposed vs Baseline}.
AM-DETR significantly outperformed the baseline in all metrics as shown in Table~\ref{tab: amr_score} (rows 4 and 6). 
Notably, the improvement of R1@0.7 and average mAP for UnAV-100 subset was 9.00 and 11.82 points.
This result suggests that capturing the temporal dependency between audio features is important for AMR.
\\
\textbf{AMR performance on SED dataset}
In addition to the Clotho-Moment evaluation split and UnAV-100 subset, we also evaluate the AMR models on TUT sound events 2017 (TUT)~\cite{Mesaros2016_EUSIPCO} used in SED.
Since the labels of TUT consist of too short words for the text query of AMR, we added several words with an average of 4.8 words to each label.
Comparing the UnAV-100 subset, the performance on TUT was significantly low.
This is because the ground-truth labels of TUT have very high and very high resolutions and contain many fine intervals. This property is different from that of Clotho-Moment, which contains few fine intervals.
\\
\textbf{Feature extractor}. We investigate the impact of contrastive learning on the feature extractor.
First, we found that contrastive learning of the feature extractor significantly improves the retrieval performance  (rows 4 and 5).
Notably, the improvement of R1@0.7 and average mAP for UnAV-100 subset was 29.00 and 23.11 points, respectively.
Next, we investigate the impact of the architecture of the feature extractors.
Comparing VR and CLAP (rows 5 and 6), the performance was comparable.
Considering that VR outperformed CLAP in audio retrieval~\cite{munakata2024dcase}, this result indicates that an encoder superior in audio retrieval is not necessarily equally good in AMR.
\\
\textbf{Window length}.
We investigated the impact of the window length of the audio encoder, and we found the shorter window length is better as shown on the left side of Figure~\ref{fig:hparam_vs_score}. 
Surprisingly, this result contrasts with the result of the baseline. as shown in Table~\ref{tab: amr_score}.
\\
\textbf{Amount of training data}.
We investigated the impact of the amount of simulated training data. 
As shown in the right side of Figure~\ref{fig:hparam_vs_score}, the performance improved with more data, and further improvement is expected as the amount of data is increased.
\\
\textbf{AMR as zero-shot SED}.
\label{sec: result}
Because the AMR models can be seen as zero-shot SED as explained in Section~\ref{sec: diff}, we evaluated AM-DETR using the same metrics as for SED.
In addition to AM-DETR, we evaluate a model that simply combines CLAP and the sliding window, which we call SED-CLAP.
The thresholds for SED were tuned with the development set to maximize the F1 score, and the threshold was set to 0.5 if it was not tuned. 
The SED performance on TUT is shown in Table~\ref{tab: sed} and we found two insights.
First, AM-DETR achieved the F1 score of over 41 points, which is 82\% of the conventional SED model based on supervised learning~\cite{Drossos2019}, despite a zero-shot condition.
Second, comparing SED-CLAP, AM-DETR is less sensitive to the threshold, which reduces tuning effort.

\begin{table}[t]
    \centering
    \caption{Performance of zero-shot SED on TUT. All metrics are measured in micro average. The window lengths of SED-CLAP and AM-DETR were 7 and 1, respectively.}
    \vspace{-3mm}
    \label{tab: sed}
    \scalebox{0.95}{
        \begin{tabular}{lccr|rrr}
        \toprule
        Method & Zero-shot & Tune & Thr. & Precision & Recall & F1 \\
        \midrule
        SED-CLAP &  $\checkmark$ & - & 0.50
        & 47.07 & 12.19 & 19.37\\
        
        SED-CLAP & $\checkmark$ & $\checkmark$ & 0.40 & 34.75 &64.94 & 45.27 \\
        
        \midrule
        
        AM-DETR & $\checkmark$ &  - & 0.50
        & 32.34 & 47.73 & 38.55  \\

        AM-DETR & $\checkmark$ & $\checkmark$ & 0.10
        & 32.66 & 55.51 & 41.12\\
        
        \midrule
        RNN~\cite{Drossos2019} & - & $\checkmark$ & N/A
        & N/A & N/A & 50.~~~ \\
        \bottomrule
        \end{tabular}
    }
    \vspace{-5mm}
\end{table}

\section{Conclusion}
We newly proposed the task of audio moment retrieval.
For training the AMR model, we proposed a simulated dataset called Clotho-Moment. 
The proposed method significantly outperformed baselines in evaluations using real-world data.
Our future work includes preparing larger-scale evaluation datasets including multiple moments relevant to a single query.
\bibliographystyle{IEEEtran}
\bibliography{mybibfile}

\begin{thebibliography}{10}
\providecommand{\url}[1]{#1}
\csname url@samestyle\endcsname
\providecommand{\newblock}{\relax}
\providecommand{\bibinfo}[2]{#2}
\providecommand{\BIBentrySTDinterwordspacing}{\spaceskip=0pt\relax}
\providecommand{\BIBentryALTinterwordstretchfactor}{4}
\providecommand{\BIBentryALTinterwordspacing}{\spaceskip=\fontdimen2\font plus
\BIBentryALTinterwordstretchfactor\fontdimen3\font minus \fontdimen4\font\relax}
\providecommand{\BIBforeignlanguage}[2]{{%
\expandafter\ifx\csname l@#1\endcsname\relax
\typeout{** WARNING: IEEEtran.bst: No hyphenation pattern has been}%
\typeout{** loaded for the language `#1'. Using the pattern for}%
\typeout{** the default language instead.}%
\else
\language=\csname l@#1\endcsname
\fi
#2}}
\providecommand{\BIBdecl}{\relax}
\BIBdecl

\bibitem{koepke2022audio}
A.~S. Koepke, A.-M. Oncescu, J.~F. Henriques, Z.~Akata, and S.~Albanie, ``Audio retrieval with natural language queries: A benchmark study,'' \emph{IEEE Transactions on Multimedia}, vol.~25, pp. 2675--2685, 2022.

\bibitem{drossos2020clotho}
K.~Drossos, S.~Lipping, and T.~Virtanen, ``Clotho: An audio captioning dataset,'' in \emph{Proc. ICASSP}, 2020, pp. 736--740.

\bibitem{kim2019audiocaps}
C.~D. Kim, B.~Kim, H.~Lee, and G.~Kim, ``Audio{C}aps: Generating captions for audios in the wild,'' in \emph{Proc. NAACL-HLT}, 2019, pp. 119--132.

\bibitem{mei2024wavcaps}
X.~Mei, C.~Meng, H.~Liu, Q.~Kong, T.~Ko, C.~Zhao, M.~D. Plumbley, Y.~Zou, and W.~Wang, ``Wavcaps: A chatgpt-assisted weakly-labelled audio captioning dataset for audio-language multimodal research,'' \emph{IEEE/ACM Transactions on Audio, Speech, and Language Processing}, 2024.

\bibitem{elizalde2024natural}
B.~Elizalde, S.~Deshmukh, and H.~Wang, ``Natural language supervision for general-purpose audio representations,'' in \emph{Proc. ICASSP}, 2024, pp. 336--340.

\bibitem{yusong2023laionclap}
Y.~Wu, K.~Chen, T.~Zhang, Y.~Hui, T.~Berg-Kirkpatrick, and S.~Dubnov, ``Large-scale contrastive language-audio pretraining with feature fusion and keyword-to-caption augmentation,'' in \emph{Proc. ICASSP}, 2023, pp. 1--5.

\bibitem{saeed2021contrastive}
A.~Saeed, D.~Grangier, and N.~Zeghidour, ``Contrastive learning of general-purpose audio representations,'' in \emph{Proc. ICASSP}, 2021, pp. 3875--3879.

\bibitem{niizumi24_interspeech}
D.~Niizumi, D.~Takeuchi, Y.~Ohishi, N.~Harada, M.~Yasuda, S.~Tsubaki, and K.~Imoto, ``{M2D-CLAP}: Masked modeling duo meets clap for learning general-purpose audio-language representation,'' in \emph{Proc. Interspeech}, 2024, pp. 57--61.

\bibitem{Primus2023}
P.~Primus, K.~Koutini, and G.~Widmer, ``Advancing natural-language based audio retrieval with passt and large audio-caption data sets,'' in \emph{Proc. DCASE Workshop}, September 2023, pp. 151--155.

\bibitem{radford2021learning}
A.~Radford, J.~W. Kim, C.~Hallacy, A.~Ramesh, G.~Goh, S.~Agarwal, G.~Sastry, A.~Askell, P.~Mishkin, J.~Clark \emph{et~al.}, ``Learning transferable visual models from natural language supervision,'' in \emph{Proc. ICML}.\hskip 1em plus 0.5em minus 0.4em\relax PMLR, 2021, pp. 8748--8763.

\bibitem{anne2017localizing}
L.~Anne~Hendricks, O.~Wang, E.~Shechtman, J.~Sivic, T.~Darrell, and B.~Russell, ``Localizing moments in video with natural language,'' in \emph{Proc. ICCV}, 2017, pp. 5803--5812.

\bibitem{gao2017tall}
J.~Gao, C.~Sun, Z.~Yang, and R.~Nevatia, ``Tall: Temporal activity localization via language query,'' in \emph{Proc. ICCV}, 2017, pp. 5267--5275.

\bibitem{lei2020tvr}
J.~Lei, L.~Yu, T.~L. Berg, and M.~Bansal, ``{TVR}: A large-scale dataset for video-subtitle moment retrieval,'' in \emph{Proc. ECCV}, 2020, pp. 447--463.

\bibitem{lei2021detecting}
J.~Lei, T.~L. Berg, and M.~Bansal, ``Detecting moments and highlights in videos via natural language queries,'' in \emph{Proc. NeurIPS}, vol.~34, 2021, pp. 11\,846--11\,858.

\bibitem{moon2023query}
W.~Moon, S.~Hyun, S.~Park, D.~Park, and J.-P. Heo, ``Query-dependent video representation for moment retrieval and highlight detection,'' in \emph{Proc. CVPR}, 2023, pp. 23\,023--23\,033.

\bibitem{moon2023correlation}
W.~Moon, S.~Hyun, S.~Lee, and J.-P. Heo, ``Correlation-guided query-dependency calibration in video representation learning for temporal grounding,'' \emph{arXiv preprint arXiv:2311.08835}, 2023.

\bibitem{xiao2024bridging}
Y.~Xiao, Z.~Luo, Y.~Liu, Y.~Ma, H.~Bian, Y.~Ji, Y.~Yang, and X.~Li, ``Bridging the gap: A unified video comprehension framework for moment retrieval and highlight detection,'' in \emph{Proc. CVPR}, 2024, pp. 18\,709--18\,719.

\bibitem{carion2020end}
N.~Carion, F.~Massa, G.~Synnaeve, N.~Usunier, A.~Kirillov, and S.~Zagoruyko, ``End-to-end object detection with transformers,'' in \emph{Proc. ECCV}, 2020, pp. 213--229.

\bibitem{cai2003highlight}
R.~Cai, L.~Lu, H.-J. Zhang, and L.-H. Cai, ``Highlight sound effects detection in audio stream,'' in \emph{Proc. ICME}, vol.~3, 2003, pp. 37--40.

\bibitem{cakir2017convolutional}
E.~Cak{\i}r, G.~Parascandolo, T.~Heittola, H.~Huttunen, and T.~Virtanen, ``Convolutional recurrent neural networks for polyphonic sound event detection,'' \emph{IEEE/ACM Transactions on Audio, Speech, and Language Processing}, vol.~25, no.~6, pp. 1291--1303, 2017.

\bibitem{komatsu2016acoustic}
T.~Komatsu, Y.~Senda, and R.~Kondo, ``Acoustic event detection based on non-negative matrix factorization with mixtures of local dictionaries and activation aggregation,'' in \emph{Proc. ICASSP}, 2016, pp. 2259--2263.

\bibitem{imoto2020sound}
K.~Imoto, N.~Tonami, Y.~Koizumi, M.~Yasuda, R.~Yamanishi, and Y.~Yamashita, ``Sound event detection by multitask learning of sound events and scenes with soft scene labels,'' in \emph{Proc. ICASSP}, 2020, pp. 621--625.

\bibitem{miyazaki2020convolution}
K.~Miyazaki, T.~Komatsu, T.~Hayashi, S.~Watanabe, T.~Toda, and K.~Takeda, ``Convolution-augmented transformer for semi-supervised sound event detection,'' in \emph{Proc. DCASE Workshop}, 2020, pp. 100--104.

\bibitem{komatsu2020scene}
T.~Komatsu, K.~Imoto, and M.~Togami, ``Scene-dependent acoustic event detection with scene conditioning and fake-scene-conditioned loss,'' in \emph{Proc. ICASSP}, 2020, pp. 646--650.

\bibitem{mesaros2021sound}
A.~Mesaros, T.~Heittola, T.~Virtanen, and M.~D. Plumbley, ``Sound event detection: A tutorial,'' \emph{IEEE Signal Processing Magazine}, vol.~38, no.~5, pp. 67--83, 2021.

\bibitem{li2023ast}
K.~Li, Y.~Song, L.-R. Dai, I.~McLoughlin, X.~Fang, and L.~Liu, ``{AST-SED}: An effective sound event detection method based on audio spectrogram transformer,'' in \emph{Proc. ICASSP}, 2023, pp. 1--5.

\bibitem{shao2024fine}
N.~Shao, X.~Li, and X.~Li, ``Fine-tune the pretrained {ATST} model for sound event detection,'' in \emph{Proc. ICASSP}, 2024, pp. 911--915.

\bibitem{liu22w_interspeech}
X.~Liu, H.~Liu, Q.~Kong, X.~Mei, J.~Zhao, Q.~Huang, M.~D. Plumbley, and W.~Wang, ``Separate what you describe: Language-queried audio source separation,'' in \emph{Proc. Interspeech}, 2022, pp. 1801--1805.

\bibitem{li2023target}
C.~Li, Y.~Qian, Z.~Chen, D.~Wang, T.~Yoshioka, S.~Liu, Y.~Qian, and M.~Zeng, ``Target sound extraction with variable cross-modality clues,'' in \emph{Proc. ICASSP}, 2023, pp. 1--5.

\bibitem{venkataramanan2023imagenet}
S.~Venkataramanan, M.~N. Rizve, J.~Carreira, Y.~M. Asano, and Y.~Avrithis, ``Is imagenet worth 1 video? learning strong image encoders from 1 long unlabelled video,'' in \emph{Proc. ICLR}, 2024.

\bibitem{fujita2019end}
Y.~Fujita, N.~Kanda, S.~Horiguchi, K.~Nagamatsu, and S.~Watanabe, ``End-to-end neural speaker diarization with permutation-free objectives,'' in \emph{Proc. Interspeech}, 2019, pp. 4300--4304.

\bibitem{geng2023dense}
T.~Geng, T.~Wang, J.~Duan, R.~Cong, and F.~Zheng, ``Dense-localizing audio-visual events in untrimmed videos: A large-scale benchmark and baseline,'' in \emph{Proc. CVPR}, 2023, pp. 22\,942--22\,951.

\bibitem{rezatofighi2019generalized}
H.~Rezatofighi, N.~Tsoi, J.~Gwak, A.~Sadeghian, I.~Reid, and S.~Savarese, ``Generalized intersection over union: A metric and a loss for bounding box regression,'' in \emph{Proc. CVPR}, 2019, pp. 658--666.

\bibitem{nishimura2024lighthouse}
T.~Nishimura, S.~Nakada, H.~Munakata, and T.~Komatsu, ``Lighthouse: A user-friendly library for reproducible video moment retrieval and highlight detection,'' \emph{arXiv preprint arXiv:2408.02901}, 2024.

\bibitem{munakata2024dcase}
H.~Munakata, T.~Nishimura, S.~Nakada, and T.~Komatsu, ``Training strategy of massive text-to-audio models and gpt-based query-augmentation,'' in \emph{DCASE Challenge, Technical Report}, 2024.

\bibitem{chen2024vast}
S.~Chen, H.~Li, Q.~Wang, Z.~Zhao, M.~Sun, X.~Zhu, and J.~Liu, ``{VAST}: A vision-audio-subtitle-text omni-modality foundation model and dataset,'' in \emph{Proc. NeurIPS}, 2024.

\bibitem{liu2019roberta}
Y.~Liu, M.~Ott, N.~Goyal, J.~Du, M.~Joshi, D.~Chen, O.~Levy, M.~Lewis, L.~Zettlemoyer, and V.~Stoyanov, ``{RoBERTa}: A robustly optimized bert pretraining approach,'' \emph{arXiv preprint arXiv:1907.11692}, 2019.

\bibitem{loshchilov2017decoupled}
I.~Loshchilov, ``Decoupled weight decay regularization,'' \emph{arXiv preprint arXiv:1711.05101}, 2017.

\bibitem{Mesaros2016_EUSIPCO}
A.~Mesaros, T.~Heittola, and T.~Virtanen, ``{TUT} database for acoustic scene classification and sound event detection,'' in \emph{Proc. EUSIPCO}, 2016.

\bibitem{Drossos2019}
K.~Drossos, S.~Gharib, P.~Magron, and T.~Virtanen, ``Language modelling for sound event detection with teacher forcing and scheduled sampling,'' in \emph{Proc. DCASE Workshop}, 2019, pp. 59--63.

\end{thebibliography}

\end{document}